\documentclass[sigconf]{acmart}

\usepackage{stfloats}
\usepackage{graphicx}
\AtBeginDocument{%
  \providecommand\BibTeX{{%
    \normalfont B\kern-0.5em{\scshape i\kern-0.25em b}\kern-0.8em\TeX}}}

\setcopyright{acmcopyright}
\copyrightyear{2022}
\acmYear{2022}
\setcopyright{rightsretained}
\acmConference[FAccT '22]{2022 ACM Conference on Fairness, Accountability, and Transparency}{June 21--24, 2022}{Seoul, Republic of Korea}
\acmBooktitle{2022 ACM Conference on Fairness, Accountability, and Transparency (FAccT '22), June 21--24, 2022, Seoul, Republic of Korea}\acmDOI{10.1145/3531146.3534637}
\acmISBN{978-1-4503-9352-2/22/06}

\usepackage{booktabs}
\usepackage{breakcites}
\usepackage{makecell}
\usepackage{natbib}
\usepackage{graphicx}
\usepackage{url}
\usepackage[utf8]{inputenc}
\usepackage{pdfpages}

\usepackage{soul}
\usepackage{float}
\def\hyphenateAndTtWholeString #1{\xHyphenate#1$\wholeString\unskip}

\def\xHyphenate#1#2\wholeString {\if#1$%
    \else\transform{#1}%
    \takeTheRest#2\ofTheString\fi}

\def\takeTheRest#1\ofTheString\fi
{\fi \xHyphenate#1\wholeString}

\def\transform#1{\url{#1}\hskip 0pt plus 1pt}

\def\urlx #1{\href{#1}{\hyphenateAndTtWholeString{#1}}}

\usepackage{verbatim}

\makeatletter
\renewcommand\noindentparagraph{\@startsection{paragraph}{4}{\z@}%
  {-.5\baselineskip \@plus -2\p@ \@minus -.2\p@\vspace{-.4em}}%
  {-3.5\p@}%
  {\ACM@NRadjust{\bfseries}}}

\renewcommand\subsubsection{\@startsection{subsubsection}{3}{\z@}%
  {-.5\baselineskip \@plus -2\p@ \@minus -.2\p@}%
  {.25\baselineskip}%
  {\ACM@NRadjust{\@subsubsecfont\@adddotafter}}}
\makeatother

\begin{document}

\title{Data Governance in the Age of Large-Scale Data-Driven Language Technology}

\author{Yacine Jernite}
\email{yacine@huggingface.co}
\orcid{0000-0002-8053-6862 }
\affiliation{
\institution{Hugging Face}
\city{Brooklyn}
\country{United States}
}

\author{Huu Nguyen}
\email{huu@ontocord.ai}
\affiliation{
\institution{Ontocord}
\city{New York}
\country{United States}
}

\author{Stella Biderman}
\email{stellabiderman@gmail.com}
\affiliation{
\institution{EleutherAI}
\city{Washington, D.C.}
\country{United States}
}

\author{Anna Rogers}
\email{anna.gld@gmail.com}
\affiliation{
\institution{University of Copenhagen}
\city{Copenhagen}
\country{Denmark}
}

\author{Maraim Masoud}
\email{maraim.elbadri@gmail.com}
\affiliation{
\institution{Independent}
\city{Dublin}
\country{Ireland}
}

\author{Valentin Danchev}
\email{val.danchev@gmail.com}
\affiliation{
\institution{University of Essex}
\city{Colchester}
\country{United Kingdom}
}

\author{Samson Tan}
\authornote{Work done prior to joining AWS.}
\email{samson.tmr@u.nus.edu}
\affiliation{
\institution{AWS AI Research \& Education}
\city{San Francisco}
\country{United States}
}

\author{Alexandra Sasha Luccioni}
\email{sasha.luccioni@huggingface.co}
\affiliation{
\institution{Hugging Face}
\city{Montréal}
\country{Canada}
}

\author{Nishant Subramani}
\email{nishant.subramani23@gmail.com}
\affiliation{
\institution{Allen Institute for Artificial Intelligence}
\city{Seattle}
\country{United States}
}

\author{Gérard Dupont}
\email{ger.dupont@gmail.com}
\affiliation{
\institution{Independent}
\city{Paris}
\country{France}
}

\author{Jesse Dodge}
\email{jessed@allenai.org}
\affiliation{
\institution{Allen Institute for Artificial Intelligence}
\city{Seattle}
\country{United States}
}

\author{Kyle Lo}
\email{kylel@allenai.org}
\affiliation{
\institution{Allen Institute for Artificial Intelligence}
\city{Seattle}
\country{United States}
}

\author{Zeerak Talat}
\email{zeerak_talat@sfu.ca}
\affiliation{
\institution{Simon Fraser University}
\city{Burnaby}
\country{Canada}
}

\author{Dragomir Radev}
\email{dragomir.radev@yale.edu}
\affiliation{
\institution{Yale University}
\city{New Haven}
\country{United States}
}

\author{Isaac Johnson}
\email{isaac@wikimedia.org}
\affiliation{
\institution{Wikimedia}
\city{Brooklyn}
\country{United States}
}

\author{Somaieh Nikpoor}
\email{smnikpoor@gmail.com}
\affiliation{
\institution{CAIDP}
\city{Toronto}
\country{Canada}
}

\author{Jörg Frohberg}
\email{jfrohb@gmail.com}
\affiliation{
\institution{apergo.ai}
\city{Leipzig}
\country{Germany}
}

\author{Aaron Gokaslan}
\email{akg87@cornell.edu}
\affiliation{
\institution{Cornell University}
\city{Ithaca}
\country{United States}
}

\author{Peter Henderson}
\email{phend@stanford.edu}
\affiliation{
\institution{Stanford University}
\city{Stanford}
\country{United States}
}

\author{Rishi Bommasani}
\email{rishibommasani@gmail.com}
\affiliation{
\institution{Stanford University}
\city{Stanford}
\country{United States}
}

\author{Margaret Mitchell}
\email{meg@huggingface.co}
\affiliation{
\institution{Hugging Face}
\city{Seattle}
\country{United States}
}

\renewcommand{\shortauthors}{Jernite et al.}



\begin{abstract}

The recent emergence and adoption of Machine Learning technology, and specifically of Large Language Models, has drawn attention to the need for systematic and transparent management of language data. 
This work proposes an approach to global language data governance that attempts to organize data management amongst stakeholders, values, and rights.
Our proposal is informed by prior work on distributed governance that accounts for human values and grounded by an international research collaboration that brings together researchers and practitioners from 60 countries.  
The framework we present is a multi-party international governance structure focused on language data, and incorporating technical and organizational tools needed to support its work.

\end{abstract}

\maketitle

\section{Introduction}

New families of algorithms relying on \textit{deep learning} have made it possible to extract ever more complex language statistics from growing numbers of text and speech records to drastically improve the performance and applicability of data-driven \textit{Natural Language Processing} (NLP) systems.
As a result, language technologies have become an integral part of daily lived experience in a greater variety of areas both online (Internet search engines, content recommendation and moderation in social media) and offline (automatic translation and speech transcription in official documents and interactions) to the point of becoming near ubiquitous.
Consequently, through being so deeply embedded into modern human life, the governance of these new forms of infrastructure \textemdash or the lack thereof \textemdash has come to exert power over individuals' and communities' lives and access to technology.

These practical applications of language technology are increasingly reliant on approaches based on trained Large Language Models (LLMs)~\citep{elmo,gpt,bert,gpt3}, whereby models are first exposed to as large and varied a collection of language data as possible with the aim of extracting \textit{``general''} properties of a language of interest.
This first step then makes it easier to fine-tune models that learn to perform a range of \textit{``specific''} NLP tasks more efficiently in that same language setting.
As such, the language corpora used to train LLMs need to meet significantly different requirements than the more purpose-specific \textit{datasets} that have hitherto supported major advances in data-driven NLP.
Indeed, while concerns of \textit{``generality''} are not new to the field of Machine Learning (ML), this two-stage approach of {(pre-)training} followed by further training and fine-tuning for a task has given them a new scope; where the properties identified by the model are expected to hold across a much greater variety of tasks, domains, and settings as long as they are in the same ``language(s)'' as the text it was pre-trained on.

\begin{figure}[t!]
    \centering
    \includegraphics[width=0.48\textwidth]{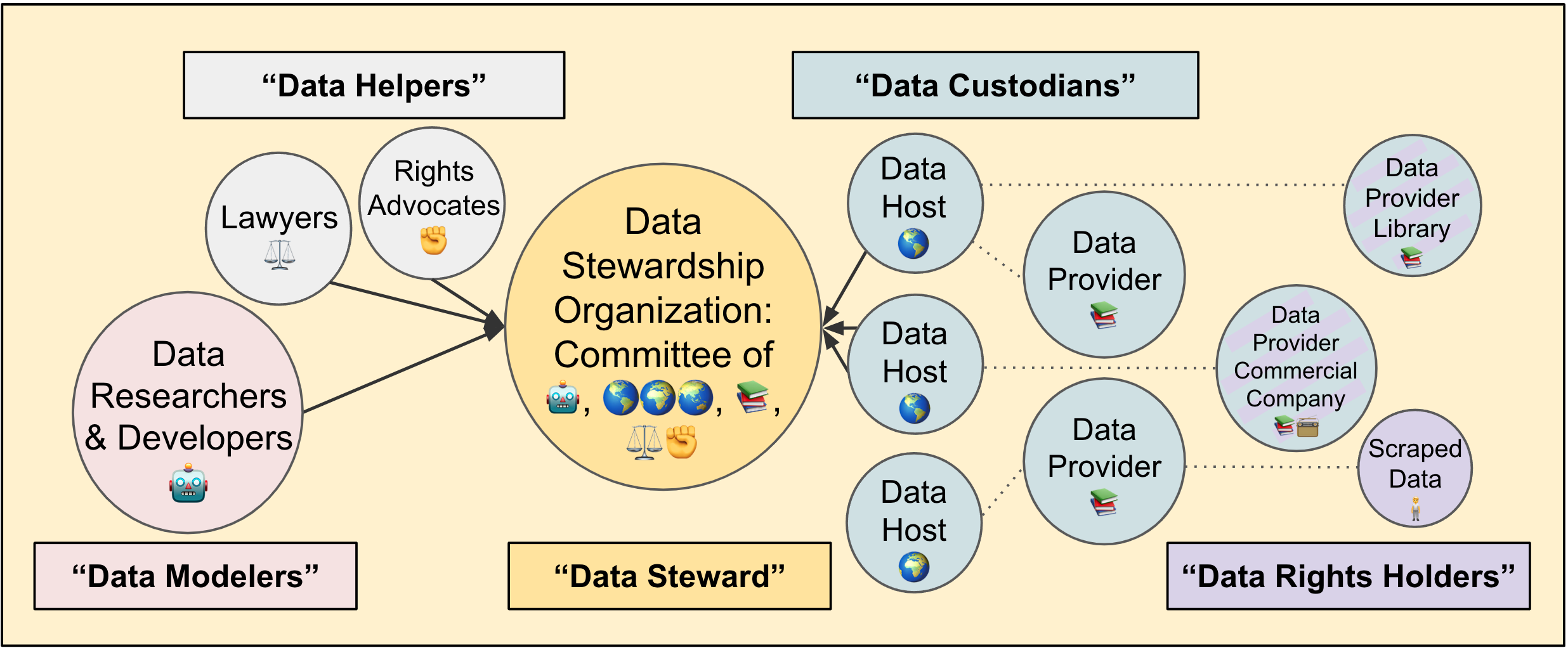}
    \caption{Overview of the Data Stewardship Organization and Actors}
    \label{fig:data-gov-org}
\end{figure}

However, whereas recent advances in modeling and hardware have increased the \textbf{data training capacity} of LLMs, increasing from Wikipedia-scale corpora to close to three orders of magnitude  more,~\footnote{The \textit{Chinchilla} model of~\citet{Hoffmann2022TrainingCL} was trained on over 1.4 trillion tokens compared to the earlier BERT's 3.3B words corpus.} devising methods for carefully identifying, obtaining, and managing a \textbf{sufficiently large and diverse collection} of language data to take full advantage of this increased capacity has remained an elusive endeavor.
Indeed, in order to support such ambitions of generality, this collection would need to include language data from a great diversity of carefully curated sources to minimize harms in downstream applications~\citep{parrots,rogers2021changing}, with international rights holders spanning multiple jurisdictions, and extend to multiple languages beyond the common English (further discussed by \citet{language_inequalities}).
This requires a more intentional approach to collecting and working with data, but designing a \textbf{data governance structure} to appropriately handle such varied data sources while respecting the \textbf{rights and interests of their stakeholders} presents a unique challenge that is only partly met by existing language data management approaches. 

To better address these needs,  we propose a new model for data governance in the form of a Data Stewardship Organization (DSO, see Figure~\ref{fig:data-gov-org} diagram) working in conjunction with related stakeholders and rights holders. The DSO primarily aims to foster the agency of data subjects and rights holders with respect to the uses of their data as the amount and diversity of contexts for this data grows.
It is designed to enable multiple stakeholders to collaborate on the decisions that go into building and managing a collection of language resources, so as to meet goals of responsible data governance at a scale and diversity that may support this new generation of data-driven language technology.
While our work is grounded by the goal of training a  multilingual LLM, we also note that many of the constraints and impacts of the design choices proposed in this work hold across a greater variety of uses of human-centric research and development data. We endeavor to also consider these related applications when relevant.

\subsection{Research Context and Paper Outline}
\label{sec:intro-context}

The research presented in this paper is conducted in the context of a year-long, distributed, collaborative workshop on Large Language Models.~\footnote{\url{https://bigscience.huggingface.co/}}
The workshop brings together over 1000 participants from 60 countries and is organized into smaller working groups focused on key aspects of the topic, including model architecture and training procedure, evaluation of performance and social biases, multilinguality, and data sourcing and governance.
Part of the value in this project is making connections between different fields of knowledge that are not normally very connected, and do so in a practical case study that pursues ethical, legal and technical goals all together.

The following sections present the findings of the data governance group in our effort to build a governance structure to manage and preserve the training data used in the project while promoting agency of all stakeholders and contending with multiple legal contexts.
Section~\ref{sec:governance} overviews prior work on the theory and mechanisms of distributed governance, outlining the special role of \textit{values} and defining the object of our governance effort.
Section~\ref{sec:context} then examines the social, legal, and technical context for using language data, and Section~\ref{sec:current-approaches} reviews current approaches to data management in ML/NLP and in Wikimedia, a distributed collaborative project whose goals and requirements have significant overlap with ours.
Finally, Section~\ref{sec:dso} describes our proposed governance structure, describing its various actors and outlining a framework for their interactions.

\section{Distributed Governance: Values and Definitions}
\label{sec:governance}

Our proposed organization aims to promote \textit{better} data governance in the context of data-driven language technology research and development.
To support this project, we start by reviewing literature on the processes and mechanisms of distributed governance (Section~\ref{sec:governance-concepts}), and in particular on the \emph{values} that underpin them (\ref{sec:governance-values}).
We then position our governance proposal with respect to these processes by defining both its object, namely human-centric data used in NLP (\ref{sec:governance-data}), and its relationship to other aspects of data management (\ref{sec:governance-scope}).

\subsection{Approaching Collaborative Governance: Theories and Mechanisms}
\label{sec:governance-concepts}

Governance is a nebulous concept, defined by the~\citet{global-governance} as \textit{``the sum of the many ways individuals and institutions, public and private, manage their common affairs''}.
Topics such as \emph{technology governance} have received increasing attention in the last few decades as the digital transformation of the late 20th and early 21st century has increased the speed at which technological innovation changes people's lives around the world~\citep{tech-governance-oecd}, leading to extensive analyses of the processes, dynamics, and particular challenges of global governance.

One such challenge has proven to be the impossibility of governing any individual subject in isolation in a fully integrated world, a phenomenon studied under the name of \textbf{regime complexes}~\citep{regime-complex}.
\citet{regime-complex-climate} study the case of the regime complex for climate change, whose global governance happens at the intersection of \textit{e.g.,} UN and local legal regimes and bilateral agreements and spans topics such as trade regulation, technology, or geoengineering. 
Consequently, governance efforts need to account for \textbf{fragmentation} when organizations in inter-connected areas make choices that have bearing on each other; by examining these connections and positioning any decision within a dense network of issues and entities~\citep{environment-governance}.
\emph{Data governance}, especially of language data, is similarly integrated in a multitude of related areas, of which Section~\ref{sec:context} will discuss the social, legal, and technical\break dimensions.

\begin{figure}
\centering
\hspace{-2em}
 \includegraphics[scale=0.1]{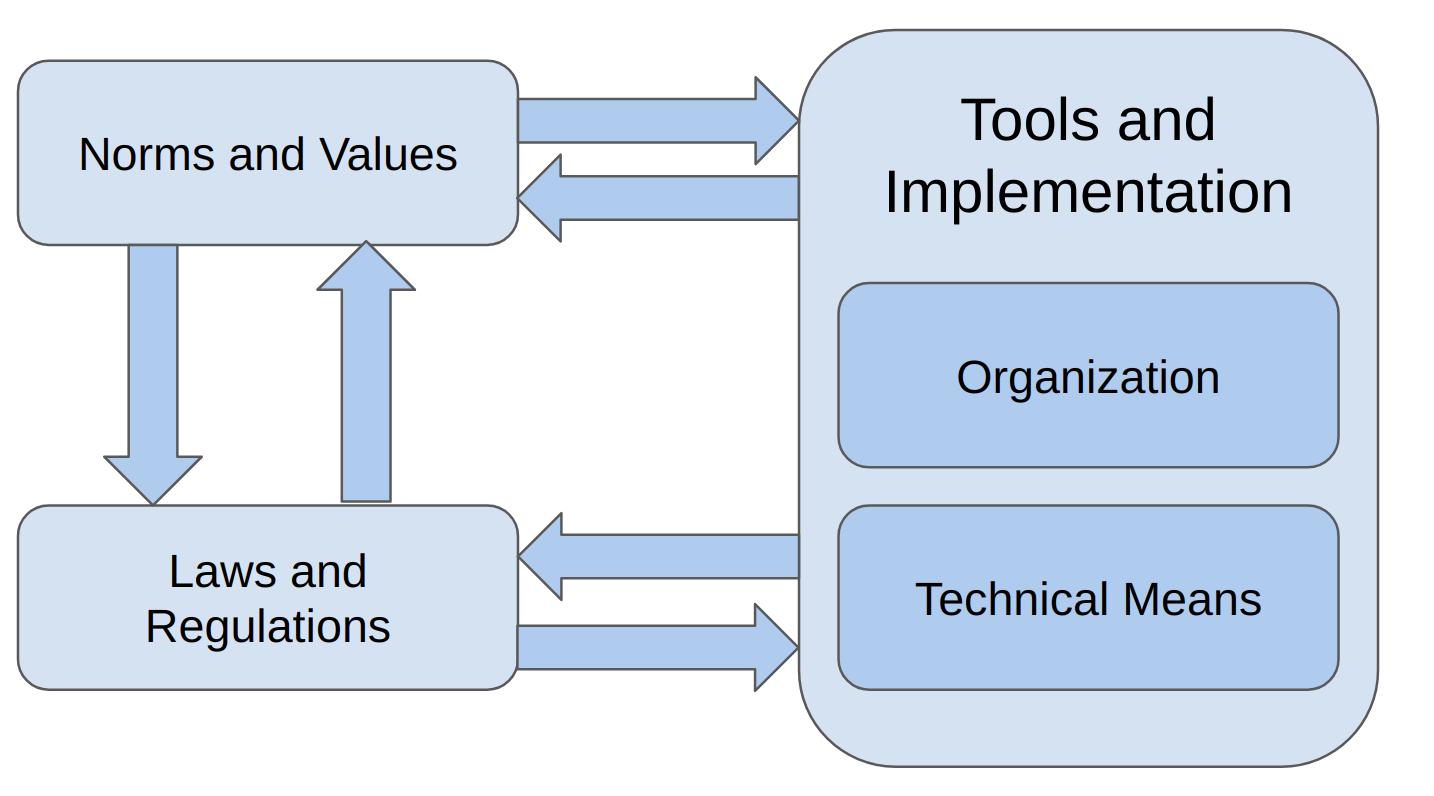}
    \caption{Collaborative governance mechanisms rely on interacting pillars.}
    \label{fig:gov-mechanisms}
\end{figure}

\begin{figure}
    \includegraphics[scale=0.3]{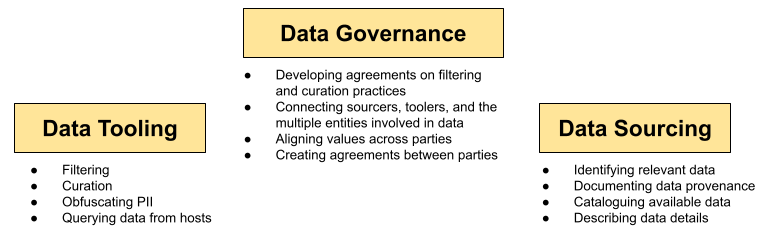}
    \caption{Machine Learning Data Triad}
    \label{fig:data-triad}
\end{figure}

Having a broad classification of the  mechanisms that underpin governance can help us better navigate this network. In addition to the \textbf{laws \& regulations} in their various regimes, previous work has focused on the specific role of \textbf{tools \& implementation} (such as indicators~\citep{Lehtonen2016TheMR} or ICT tools \citep{Pereira2003ICTTT}) and on the importance of \textbf{norms \& values}~\citep{Kooiman2009METAGOVERNANCEVN} in governance efforts.
In general, we can map mechanisms reviewed in governance literature to one or to the intersection of two of these pillars (Figure~\ref{fig:gov-mechanisms}).
For example, \citet{algorithm-accountability}'s aspects of \textit{algorithmic accountability} include mechanisms such as \textit{prohibitions and moratoria} (\ul{regulations}), \textit{principles and guidelines} (\ul{norms}), or \textit{independent oversight bodies} (organizational \ul{tools}).
A similar analysis may be applied to works studying governance's aim to \textbf{identify and resolve tensions} between actors.
\citet{interactive-collab-governance}'s proposed framework considers \textit{principled engagement} and \textit{shared motivation} between all the participants in a governance structure (their shared \ul{values}) as the basis for resolving tensions.
The approach of \citet{institutional-collective-action} addresses dilemmas stemming from different externalities, such as different local \underline{regulations} of the object of governance.
\citet{tech-ecosystem-governance} examine the case of governance of software platforms (specifically the \underline{tools} they rely on) through the lens of striking a balance between a system's stability and ability to evolve.
In order to position our own governance efforts with respect to all these processes, we review its \ul{values} in Section~\ref{sec:governance-values}, relations to technical \ul{tools} in \ref{sec:governance-scope} and to \ul{regulations} in~\ref{sec:context-legal}.

Finally, previous work has also pointed out how the very mechanisms used to resolve tensions can \textbf{shift or entrench power imbalances}, and advised to pay special attention to this phenomenon.
\citet{power-governance} and \citet{framework-power-assess} examine how authority, resources, and discursive legitimacy can lead to exclusion within collaborative governance efforts.
In particular, \citet{Mohamed2020DecolonialAD} call attention to the ``first-mover advantage'' phenomenon in setting standards in the contact of AI governance: values that protect and are of interest to the people who write the standards will necessarily be prioritized over values whose upholding is more urgently needed by other parties.
We endeavor to be cognizant of these risks in our own governance proposal, both in the expression of its driving values (Section~\ref{sec:governance-values}) and of its structure and processes (Section~\ref{sec:dso}).

\subsection{Values of Governance, Governance of Values}
\label{sec:governance-values}

Section \ref{sec:governance-concepts} identifies \textit{\underline{norms}} and \textit{\underline{values}} as main pillars of governance, which are implicitly or explicitly defined by the organizations contributing the the governance structure~\citep{Kooiman2009METAGOVERNANCEVN}.
These shape design choices and trade-offs, and a static set of values, or ones expressed exclusively by the originators of the project, can lead to exclusion~\citep{power-governance} and reinforce disparities~\citep{Mohamed2020DecolonialAD}.
In this context, taking time to examine the values driving our own project, the framework that is used to contextualize them, and the way they themselves are governed
is particularly important.

\begin{table*}
    \centering
    \small
    \begin{tabular}{p{11pc}|p{15pc}}
    \toprule 
    \textbf{Value} & \textbf{Description} \\
    \midrule
    \hline 
    {\sc Inclusion, Representation,  \& \sc Non-Discrimination} &
    {Equal access to cultural resources and ability to interact with language infrastructures and technology without prejudice} \\
    \hline 
    {\sc Autonomy, Consent \& \sc Contestation} &
    {Right of individuals and communities to meaningfully control the inclusion of  their language data in public resources} \\
    \hline 
    {\sc Privacy} &
    {Right of individuals to control who may have access to their personal  identifying information (PII)} \\
    \hline 
    {\sc Just Rewards} &
    {Right to share in the financial and social benefits stemming from uses of an individual's or communities' language and data} \\
    \hline 
    {\sc Licensing \& Attribution} &
    {Right to legal controls over one's data and the product of one's work} \\
    \hline
    \hline
    {\sc Local Knowledge} &
{Local expressions of values and their context take precedence when  making and implementing local decisions} \\
    \hline
    {\sc Participation} &
    Above values and definitions evolve based on actors' needs and feedback \\
    \hline
    {\sc Beneficence} &
    Above values subject to a general ``do no harm'' approach \\
    \bottomrule
    \end{tabular}
    \caption{Set of values proposed to guide our data governance effort. }
\label{tab:bigscience_initial_values}
\vspace*{-10pt}
\end{table*}

\citet{birhane2021values} review recent literature in ML to identify values that are typically put forward to motivate work in that field.
They note that most of these focus on endogenous notions of technical performance and novelty, and leave out considerations of broader context and impact that are necessary to shaping a governance effort.
Inspired by their approach, we reviewed the working documents of the LLM workshop grounding this paper (see Section~\ref{sec:intro-context})\footnote{Records organized by working groups are publicly available at~\url{https://drive.google.com/drive/folders/1db2hYZuRs2VjoIrVaVtZJ5FLE2iS7z3p}.\\ Appendix~\ref{app:values} describes the interactions that led to the initial set of values in more details.}, and found that notions of \textbf{inclusivity} and \textbf{non-discrimination} regularly appeared.
Many of the participants' comments were also informed by the recent European drive towards more data and algorithmic regulation, including their focus on \textbf{respecting privacy} and promoting the \textbf{agency of data and algorithm subjects}\footnote{\url{https://eur-lex.europa.eu/eli/reg/2016/679/oj}}.
Additionally, and in reaction to recent practices of indiscriminate use of crawled web text, participants expressed a concern for \textbf{respecting rights} of the text creators (e.g., copyright or intellectual property laws).
Finally, participants, especially ones with ties to Africa\footnote{\url{https://www.masakhane.io/}} and South-East Asia, pointed out the potential for exploitative data practices in fully open ML data and research~\citep{Minasny2020GlobalSS}, and stressed the need for \textbf{equitable distribution of the benefits} stemming from data use and work.

Among these values, the stated goal of \ul{inclusivity} merits further examination.
Our project aims to govern global language data, which as we shall see in Section~\ref{sec:context-social} shows significant variation across cultural and social contexts.
Meanwhile, the participants of our research project remain embodied in their own subjectivities~\citep{Harawy1988SituatedKT}, which necessarily represent but a small portion of these contexts.
As such, devising a governance structure based solely on values expressed by our participants runs the risk of prioritizing their interests and excluding visions that may be more relevant to other language users~\citep{Greene2019BetterNC}.
Additionally grounding the definition of our proposed values in \textit{Human Rights frameworks} constitutes an appealing starting point to addressing these limitations given their global reach, varied realizations (both historically and geographically), and general recognition as an accepted foundation of good governance~\citep{ai-principles-consensus}.
Indeed, we find that documents such as the UDHR~\citep{united1949universal}, ICCPR~\citep{assembly1966international_iccpr}, or ICESCR~\citep{assembly1966international_icescr} echo the proposed values of \ul{non-discrimination}, \ul{privacy} and \ul{just rewards} respectively, and help ground them in an external system \cite{PrabhakaranEtAlHumanRights}. 

At the same time, while the principle of Human Rights does have a universal scope, the staggering number and diversity of human rights documents written both at the UN~\footnote{\url{https://www.ohchr.org/en/professionalinterest/pages/coreinstruments.aspx}} and regional level brings to light the inadequacy of focusing on a limited set of human rights document as absolute grounding when outlining values that apply \textit{equitably} to a global and contemporary setting, as it arises from significant differences in their philosophical foundations across the globe~\citep{roux2013human}. 
Scholarship at the intersection of decoloniality and human rights in particular has called out the need to question the universality of how we conceptualize the ``human'' in Human Rights~\citep{Fanon1952BlackSW,coloniality-hr}.
Furthermore, the focus of human rights discourse has historically been on the relationship between the individual and the state~\citep{Ratner2001CorporationsAH}, whereas the language data we propose to govern is created and managed at various meeting points between individuals, communities, corporations, and state organizations~\citep{viljoen2020democratic}.
Thus, acknowledging both the need for respecting local expertise and conceptualization of human rights~\citep{Adami2014HumanRF} and for more \emph{relational and community-oriented} notions of justice~\citep{masakhane-participatory,relational-ethics}, we complement our initial set of values as outlined in Table~\ref{tab:bigscience_initial_values} to explicitly include the prioritization of \textbf{local knowledge} when realizing shared aspirations in the context of participants' own expression of values, and conversely \textbf{participation} in shaping these over-arching governance values to better account for their specific needs.

\subsection{Object of Governance: Distributed Language Data} 
\label{sec:governance-data}

We apply the norms and values described above to a governance structure for language data, paying special attention to its use in ML datasets.
In the ML world, \ul{data} refers to any digital representation of acts, facts or other information in forms that include text, images, video, or audio recordings \cite{EuropeanDataGovernanceAct,viljoen2020democratic,PaulladaEtAl2021}, which may be collated or formed into more complex information \cite{MontrealDataLicense}.  It is often described via analogy as food and fuel for ML systems, water and oil as an AI resource, and increasingly as records of human activity \cite{StarkHoffman2019}. As such, it is a fundamental catalyst for the creation of artificial intelligence systems \cite{MontrealDataLicense}, used for training, developing and/or testing AI systems in the form of \ul{datasets}, an organized collection of data for a defined task \cite{PaulladaEtAl2021}.

The proposed data governance structure introduced in this paper focuses on \textbf{digital language data}, which includes text from news and academic articles, reports, white papers, blogs, social media posts, radio shows, and digitized books. All such data is created by a person, group of people, or an organization who may hold the rights to that data. All of these dimensions of the language data and datasets it's organized as have bearing on the governance choices (see Table~\ref{tab:data_categories} for examples of categories).
Of particular sensitivity in data governance is \textbf{human-centric data}, data that additionally refers to or represents the ideas of a person. Some kinds of textual data such as weather reports are less likely to harm an individual in case of lax governance or misuse, but human-centric data brings with it concerns around how an individual is represented, how that representation may affect them, the represented individual's consent, and other fundamental and legal rights \cite{EuropeanPersonalData}.
The specific risks to be considered depend on the individual's relationship to the data, from creators who may have commercial rights on the product of their work to users and (passive) subjects of the technology developed based on the data.
Table~\ref{tab:data_stakeholders} lists these various stakeholders, and we explore their different needs and interactions with roles in our proposed structure in Section~\ref{sec:dso}.

\begin{table}
\small
\begin{tabular}[t]{@{}l|p{4cm}}
{\bf Category}& \textbf{Examples} \\\hline
     \textit{Domain} &  news, medical, legal\\
     \textit{Genre}& literature, social media, articles\\
     \textit{Legal status} & public use, non-commercial\\
     \textit{Origin} &person, organization\\
     \textit{Source}&book, social media platform, radio\\
     \textit{Modality}&text, audio, video\\
     \textit{Goals} & curated corpus, benchmark, convenience sample\\
\end{tabular}
\caption{Dimensions of digital language datasets.}\label{tab:data_categories}
\vspace*{-10pt}
\end{table}

\begin{table}
\small
\begin{tabular}[t]{@{}l|p{4cm}}
\textbf{Stakeholders }&\textbf{ Examples} \\\hline
\textit{Data subjects} & people(s) being talked to/about\\
\textit{Data creators} & journalists, social media users \\
 \textit{Data aggregators} & social media platforms \\
\textit{Dataset creators}& researchers, organizations\\
\textit{Dataset distributors} & researcher, university, dataset hub\\
\textit{Dataset users}& model developers\\
\textit{Those affected} & users/subjects of ML systems
\end{tabular}
\caption{Stakeholders of human-centric data.}
\label{tab:data_stakeholders}
\vspace*{-10pt}
\end{table}

\subsection{Focus of a Data Governance Organization} 
\label{sec:governance-scope}

The life cycle of the data and datasets we aim to govern spans many different stages, including: data creation, selection, curation, documentation, dissemination, hosting and serving, conservation, tracking, versioning, and deletion \cite{HutchinsonEtAlDatasetAccountability,PaulladaEtAl2021}.
Management of each of these different stages will impact our ability to support the values outlined in Table~\ref{tab:bigscience_initial_values}, which will also depend on the characteristics of the data along the dimensions illustrated in Table~\ref{tab:data_categories} and on which stakeholders are most directly involved (Table~\ref{tab:data_stakeholders}).
In order to better define the specific scope of our governance proposal in all of the many combinations of these parameters, we need to be able to differentiate what falls within its purview from what may better be addressed by other aspects of data management at these various stages.

Specifically, our approach to data governance separates work done with the data, such as selection and curation, from work done around data access, control, and exchanges between different data actors.
Our focus may be seen as \emph{people-centric}, narrowing in on the people represented and the users of the data, rather than on its analytics.
Thus, we make a distinction between \textbf{Data Governance}, \textbf{Data Sourcing}, and \textbf{Data Tooling} as outlined in Figure~\ref{fig:data-triad}. 
These three directions are complementary: \ul{Data Governance} provides an overall structure wherein \ul{Sourcing} and \ul{Tooling} can come into play.
The governance work provides norms, frameworks, and communication mechanisms in order to \textit{e.g.} help operationalize definitions of \ul{contestation} in different \ul{legal contexts} to allow for the development of \textit{locally relevant} supporting \ul{tools}, or to formalize relations between actors in different roles and parts of the world.
We illustrate this categorization further on three concrete aspects of data management next.\vspace*{6pt}

\noindent\textbf{Data Governance supports Data Sourcing.}
While governance focuses on data stakeholders and interacting norms, values, laws, etc., the governance structure operates over the datasets provided through \ul{data sourcing} efforts, which accumulate, categorize, organize, and document data for datasets.
Governance provides a framework for helping \ul{sourcing} actors formalize rules on how the data they propose may be used and processing requirements.
These frameworks are designed to enable values such as \ul{representation} by e.g. removing barriers to entry and lowering risk of participation in technology to enable actors with \ul{local knowledge} to identify and fill gaps in available language resources within the global organization~\citep{BigScienceCatalogue}; however the diversity of the sources represented in the governance structure will, ultimately, depend on the quality of the \ul{sourcing} efforts.\vspace*{3pt}

\noindent\textbf{Handling Personal Information and Privacy.}
Table~\ref{tab:bigscience_initial_values} includes values of \ul{privacy} and \ul{consent}. To uphold these we need to understand what constitutes and be able to identify instances of Personal Identifiable Information (PII; a term used in the U.S.) and Personal Data (a term used in the U.K., E.U., and some other jurisdictions). All three directions illustrated in Figure~\ref{fig:data-triad} come into play for this aspect: (a) \ul{governance} helps guide the focus on the relevant aspects of personal information depending on the data types (Table~\ref{tab:data_categories}), with local legal context shaping policies for what to do with that information (e.g. whether it is indexed, obfuscated, accessible); (b) \ul{tooling} implements these definitions into software that can look for instances of personal information at scale in large amounts of text data; (c) \ul{sourcing} makes decisions on what data to prioritize based on the identified privacy risks and impact on various stakeholders.\vspace*{3pt}

\noindent\textbf{Contestation and Removal of Data.}
We also want our proposed structure to promote \ul{contestation} rights and control over one's data, in particular by allowing parties who have personal or commercial rights on data included in the organization to request its removal. This aspect also exemplifies the interaction between \ul{governance} and \ul{tooling} responsibilities. The former defines actionable guidelines and processes for identifying what consitutes a \textit{legitimate removal request} depending on the local norms and regulations of the requester and data custodian. The latter needs to ensure that the data instance can be easily found and deleted in datasets. In particular, deletion can only be meaningfully enacted if the governance structure ensures \textit{non-dissemination of the data}; that is, if data modelers and researchers can use it without making and broadcasting their own copies. This needs to rely on a combination of technical tools and, when they aren't mature enough, signed agreements or licenses defining the parameters of their data access. We describe a framework for these agreements in Section~\ref{sec:dso}.

\section{Language Data: Social, Legal, and Technical Context}
\label{sec:context}

Section~\ref{sec:governance} outlines the general mechanism of governance (\ref{sec:governance-concepts}), the values supporting our effort (section~\ref{sec:governance-values}), the kind of human-centric data we focus on (section~\ref{sec:governance-data}), and its relationship to other aspects of data manangement (\ref{sec:governance-scope}).
Our next step is to investigate the interplay between the object of our governance effort and its broader context: the social context of language data (section~\ref{sec:context-social}), the relevant legal principles and frameworks (section~\ref{sec:context-legal}), and the culture around language data use in ML and NLP (section~\ref{sec:context-technical}).

\subsection{Social Context: Social Variation and Language Discrimination}
\label{sec:context-social}

The governance values outlined in Table~\ref{tab:bigscience_initial_values} include \ul{inclusion \& non-discrimination}.
Section~\ref{sec:governance-concepts} also cautions against the risk that governance mechanisms might entrench inequalities and power disparities if those are not explicitly taken into account.
In order to apply these values and avoid those risks in the context of language data, we need to consider its social dimension. 
Here, we review  sociolinguistics literature to identify social variables that can engender discrimination in inter-personal interactions and meetings with technology.

Most named human languages are collections of language varieties with differences that stem from demographic factors such as education, geography, race, and socio-economic class \citep{labov1994principles}. However, there is a common misunderstanding that there are well-defined boundaries between languages, each with only a single grammar, lexicon, and orthography, and this has resulted in the stigmatization of the language varieties not associated with status and power \citep{kernan1971language,giron1982chicano,rahman2008middle,garcia2015sociolinguistic}, negatively impacting speakers's access to social infrastructures (\textit{e.g.} schools~\citep{cremona1977development,hasan2009semantic}, courtrooms~\cite{ling-on-trial}).

This misunderstanding permeates in modern NLP practices. For instance, texts which display sociolinguistic variation, e.g., social media text, are often labelled as ``noisy'', while text from prestige variants are deemed ``clean''. Such \emph{politics of dirt} \citep{Douglas_1978} reveal attitudes that stigmatize minority language variants \citep{giron1982chicano,rahman2008middle,garcia2015sociolinguistic} (as well as demeaning the people that speak them) whilst obscuring values and information signaled through dialectal use of language \citep{Rahman_2012}. 
``Clean'' text additionally has  been misrepresented as being ``unbiased'' against any community---a notion that has been strongly contested~\cite{bolukbasi2016man,hutchinson2020social,Talat_disembodied_2021}.
Unsurprisingly, gendered and racial disparities have been documented in a number of language technologies \citep{racial-asr,xu-etal-2021-bot,dathathri2019plug}, and processes of creating resources and technologies may further entrench such disparities \citep{zhao-etal-2017-men,Davidson_racial_2019,cao&daume2021}. For more detail see \citep{field2021survey}.

While social and linguistic discrimination do not originate or end in language technologies, such technologies do engage in society as sociotechnical systems that are imbued with values \citep{birhane2021values},
and it is therefore important to consider their role in discrimination, and the ways in which values of \ul{non-discrimination} can be implemented when governing data.
To this effect, we should be cognizant of the existing linguistic discrimination present in our societies \citep{ling-on-trial} and be careful not to inadvertently replicate them \citep{tan-etal-2020-morphin,Thylstrup_Talat_2020}.
In the context of the various interactions of \ul{data governance}, \ul{sourcing}, and \ul{tooling} mentioned in Section~\ref{sec:governance-scope}, this requires prioritizing the needs of currently under-represented language communities in their sourcing efforts, promoting notions of data quality that do not confound noise with sociolinguistic variation \citep{tan2022thesis}, and explicitly including and giving a say in the various governance choices to speakers of all language variants. The role of these data advocates is further outlined in Section~\ref{sec:dso}.

\subsection{Legal Context: Rights and Regulations}
\label{sec:context-legal}

Figure~\ref{fig:gov-mechanisms} presents \ul{laws and regulations} as one of the pillars of governance. In particular, the notion of \textbf{protected rights} can help us understand how the guiding values presented in Table~\ref{tab:bigscience_initial_values} are understood and regulated in various legal contexts. The global landscape of relevant laws is vast, but in this section we provide a brief overview of how the values of \ul{just rewards}, \ul{attribution}, and \ul{contestation} are related to the \textbf{property rights}, \ul{consent} and \ul{privacy} - to \textbf{privacy rights}, and \ul{non-discrimination} - to \textbf{user rights}.

First, we examine \textbf{property rights} for language data creators. In the U.S., property rights are often thought of as a ``bundle of sticks''~\citep{grey20142}. That is, property rights are composed of different types of rights: the right to profit from the property (i.e., receive \ul{just rewards}), to require proper crediting for re-use (i.e., \ul{attribution}), etc.
For example, an artist is entitled to fully profit from their work, or to remove it from circulation at any time.
This bundle of rights comes with some common limitations.
Copyright, trademarks, and patents can expire and ``fair use'' exemptions to copyright exist to allow certain uses of copyrighted data deemed socially beneficial, such as  keeping content from disappearing \citep{lemley2020disappearing}.
In the context of ML, there is an ongoing debate about whether and when using copyrighted data for training models constitutes ``fair use''~\citep{lemley2020fair}.
The U.S. Copyright Office recently issued an exemption to liability for removing digital rights management software for the purposes of text and data mining for non-commercial research.\footnote{\url{https://public-inspection.federalregister.gov/2021-23311.pdf}} Japan and Europe have passed similar legislation making it easier to use data for text and data mining for research purposes\footnote{\url{https://eare.eu/japan-amends-tdm-exception-copyright/}} {}\footnote{\urlx{https://www.europarl.europa.eu/news/en/press-room/20190321IPR32110/european-parliament-approves-new-copyright-rules-for-the-internet}}; this tension between social benefits from allowing re-use of data and the social harms to data creators has led some in the U.S. to call this a ``fair use crisis''~\citep{sobel2017artificial}.
The role of a governance structure will be to help data creators, hosts, and modelers navigate these tensions by providing locally relevant frameworks for \ul{contestation} and use case restrictions.

Second, we examine \textbf{privacy rights} of data creators. The view of privacy protection based on data as property ~\citep{ritter2017regulating,stepanov2020introducing,jurcys2020ownership} has been criticized as placing a substantial burden on the free flow of information, while potentially not improving privacy protections~\citep{kerry2019data}. An alternative approach to \ul{privacy} is restricting the processing of ``personal data'', as it is done in the E.U.'s General Data Protection Regulation (GDPR)~\citep{EUdataregulations2018}. This approach hinges on defining what is ``personal'', and how that interacts with ``publicly available''. For digital language data, a big issue is that most longer texts are unique and difficult to anonymize~\citep{chaski2013best,kumar2017army,mozes2021no}, and by themselves could identify people: e.g. a simple search could identify the author of a tweet, who willingly made the authorship information public. Since that author may not be aware of the visibility of their language data and conceptions of the social benefits (or lack thereof) of various research practices~\cite{Fiesler2018ParticipantPO}, privacy legislation such as GDPR may require that data be used with \textit{(revocable) \ul{consent}}, and for the \textit{specific purposes} that are clearly explained to the data subjects, who should also have the right to delete or rectify existing records (so as to enable e.g. factual corrections, updates to the previously accurate records, or the `right to be forgotten'). Given that trained ML models might be queried for specific information about individuals~\cite{carlini2020extracting}, a governance model would have to consider not only whether and how to remove specific instances from its datasets, but also how to minimize the risk of memorization when sharing the data for model training and development.

Third, we examine \textbf{user rights}. Depending on the jurisdiction, there may be an orthogonal set of laws that aims to ensure the rights of the \textit{users of models created from the data}. A number of prior works, particularly from a U.S.-centric perspective, have connected ML to legal frameworks for human rights, especially anti-discrimination~\citep{yu2018intellectual,engstrom2020algorithmic,grace2019machine,huq2019constitutional,zuiderveen2020strengthening,ho2020affirmative,engstrom2021disparate,xiang2021reconciling}. 
These works often focus on the difficulties of constraining algorithmic \ul{discrimination} in many contexts, proposing alternative legal frameworks that would allow for more regulatory enforcement of algorithmic bias. 
The evolving nature of human rights law, civil rights law, and ML may place more constraints on data curators to ensure that downstream models are more fair -- and respect rights like equal protection, anti-discrimination, or constraints on arbitrary enforcement. 
For example, New York City now regulates automated employment algorithms and would require yearly bias audits.\footnote{Administrative Code of the City of New York, Title 20, Section 1, Chapter 5, Subchapter 25.}
However, the effectiveness of these relatively new laws has yet to be tested, and in the past governments themselves have tried to leverage ML systems in potential violations of human rights.~\footnote{\url{https://notechforice.com/wp-content/uploads/2021/10/Deadly.Digital.Border.Wall_.pdf}}~\footnote{\url{https://www.reuters.com/world/china/china-uses-ai-software-improve-its-surveillance-capabilities-2022-04-08/}}  We refer the reader to cited works for more in-depth analysis of these issues, including: accessibility rights,
\footnote{For example, the Americans with Disabilities Act of 1990 (ADA) 42 U.S.C. \S\S 12101-12213, in the U.S. enabled the National Association of the Deaf to argue that automated captions in some cases were of such unacceptable quality that they did not satisfy the accessibility rights of deaf data users. \emph{See, e.g., National Ass'n of the Deaf v. Harvard University}, 377 F. Supp. 3d 49 (D. Mass. 2019); \emph{National Ass'n of the Deaf v. Netflix, Inc.}, 869 F. Supp. 2d 196 (D. Mass. 2012).}
a right to explanation,
\footnote{GDPR \cite{EUdataregulations2018} does not explicitly guarantee it, but it it does require the data processor to provide `meaningful information about the logic involved' in fully automated decisions, which could be interpreted that way \cite{SelbstPowles2017meaningful}. In May 2021 a Dutch court upheld this principle for the first time: a ridesharing company was obliged to ``communicate the main assessment criteria and their role in the automated decision [to the drivers], so that they can understand the criteria on the basis of which the decisions were taken and they are able to check the correctness and lawfulness of the data processing'' \cite{Peers_2021_EU_Law_Analysis_Ola_Uber_judgments}.}
and a right to a certain level of performance.
\footnote{The current proposal for the EU AI Act \cite{2021_AIA} distinguishes between application areas on the basis of risk they pose, and would institute external ``conformity assessments'' for the more risky applications.}
Data governance supports these user rights by allowing marginalized populations better control over how they are represented in the data used to train ML systems in an effort to lessen algorithmic discrimination, and by supporting auditability of these systems to promote accountability~\cite{algorithm-accountability}.

\subsection{Machine Learning Context: Challenges and Incentives}
\label{sec:context-technical}

One of the major challenges in creating a data governance structure for ML datasets lies in the limited amount of research on this subject within the ML community.
Very recent research --- most published within the last year --- has begun to analyze dataset values \cite{birhane2021values,denton2021genealogy}, question assumptions around dataset use \cite{PaulladaEtAl2021}, unpack what is represented in ML datasets \cite{DodgeSapEtAl_2021_Documenting_English_Colossal_Clean_Crawled_Corpus,scheuerman2021datasets,luccioni2021s} and establish basics of how an organized dataset lifecycle might proceed \cite{HutchinsonEtAlDatasetAccountability}. These just begin to scratch the surface of what well-defined data systems may look like in ML.

We see several reasons for the limited attention to data governance in ML to date. First, the mainstream ML research focuses predominantly on improvements to the model architecture, training procedure, and (hyper)parameters~\citep{sambasivan2021everyone,PaulladaEtAl2021}. For LLMs in particular, the data used to train them are one further step removed from the task-specific models built from them, so the link between data and ML progress is even more abstracted~\cite{Raji-et-al-2021,Liao-et-al-2021}. Second, research addressing dataset choices, creation, and curation, is systematically ``under-valued and de-glamorised'' \citep{sambasivan2021everyone,reviews-values-paper}\footnote{For a direct example of how the ML community treats work on datasets and values, see reviews for \cite{reviews-values-paper} \href{https://twitter.com/willie_agnew/status/1465756111676469250}{here}}. 
Even works that do include significant curation efforts for the sake of improving models~\cite{pile,raffel2019exploring} focus on definitions of quality that prioritize technical performance over the agency of data and algorithm subjects, which can result in widespread data that proliferates misogyny, pornography without consent, and malignant stereotypes \cite{BirhanePrabhuKahembwe2021}.

One approach put forward in recent years to foster more accountability of these data practices has been documentation standards for data and models in natural language processing \citep{BenderFriedman_2018_Data_Statements_for_Natural_Language_Processing_Toward_Mitigating_System_Bias_and_Enabling_Better_Science,GebruMorgensternEtAl_2020_Datasheets_for_Datasets} and ML in general \citep{MitchellWuEtAl_2019_Model_Cards_for_Model_Reporting}.
There has also been an increased focus on analyzing other dimensions of data quality and stewardship~\citep{stewardship-lessons,sambasivan2021everyone,prabhu2020large,rogers2021just}, with several noteworthy initiatives aiming to document both existing \citep{retrospective-bookcorpus,Birhane2021MultimodalDM,DodgeSapEtAl_2021_Documenting_English_Colossal_Clean_Crawled_Corpus,CaswellKreutzerEtAl_2021_Quality_at_Glance_Audit_of_Web-Crawled_Multilingual_Datasets}, and newly developed \cite{gehrmann2021gem,wang2021adversarial,pile-datasheet} resources.
These efforts have gone hand-in-hand with efforts centered around values of \emph{transparency and replicability} in scientific work, through the introduction of standards and conference checklists ~\cite{dodge2019work,pineau2021improving}.~\footnote{mostly geared towards code and experiment tracking, but also covering training and evaluation data}
The two directions have come together in the last year to extend the approach beyond simply reproducibility, with newer checklists for ``Responsible NLP''~\citep{NeurIPS_2021_Paper_Checklist,rogers2021just,ARR_2021_Responsible_NLP_research_Checklist} asserting the importance of respecting values including \ul{non-discrimination} (fairness), \ul{consent}, or \ul{privacy} in the development and use of datasets and encouraging intentional handling of data tools (see Section \ref{sec:governance-scope}).
Given the importance of conferences in the field, we may hope that these paper checklists will have a significant role to play in spreading \ul{norms} and best practices of data curation and documentation.
Still, within this context, comparatively little attention is paid to the later stages of the data life cycle (see Section~\ref{sec:governance-data}), or to data management models that intentionally include data subjects.
We review common approaches to hosting and distributing ML data in Section~\ref{sec:current-approaches}.

\section{Efforts and Challenges in ML Data Governance}
 \label{sec:current-approaches}

Despite its unquestionable importance in contributing towards higher-quality LLMs and stakeholder agency, explicit data \emph{governance} remains a relatively new field of practice in the ML and NLP communities. In this Section, we first survey existing data management efforts in AI, then provide a short description of the data governance practices in the Wikimedia project (Section~\ref{sec:current-approaches-wikimedia}), an example of a governance framework with goals and priorities similar to ours.

\vspace*{3pt}

\noindent\textbf{Centralized Dataset Management.}
Perhaps the most common method for managing NLP datasets is for the developers themselves to host the data upon release on platforms such as GitHub and personal websites. Commonly used larger organizations include \href{https://msropendata.com/}{Microsoft Research Open Data} and \href{https://allenai.org/data}{Allen Institute for AI Datasets}, as well as consortia such as the \href{https://www.ldc.upenn.edu/}{Linguistic Data Consortium} (LDC), \href{http://catalogue.elra.info/en-us/}{European Language Resources Association} (ELRA) and CLARIN~\citep{Kelli2018ImplementationOA}, which aim to centralize and standardize access to textual resources for members of the community.
An advantage of such centralized repositories is that members can access a wide range of datasets that persist unchanged over time. 
For example, any researcher who downloads the popular \href{https://catalog.ldc.upenn.edu/topten}{OntoNotes} will have the same version as other researchers, enabling reproducibility and fair comparisons. 
There are also downsides (such as membership cost or time lags), but critically for this work,  there is no place for multiple stakeholders and rights-holders to align on priorities, giving the governing organization full say over how the data should be shared and used. Data subjects and providers generally do not have visibility into the data decisions, nor recourse to address how they are represented, and centralized decisions regarding content do not necessarily account for knowledge local to where the data instance is sourced.\vspace*{3pt}

\noindent\textbf{Public Dataset Repositories.}
In recent years public repositories of datasets, like the UCI ML Repository \cite{Dua:2019} and the Hugging Face Dataset Repository \cite{lhoest-etal-2021-datasets}, have become popular.
These repositories resemble centralized dataset management, but rely primarily on user contributions, both to source datasets and to govern them.
For example, dataset submitters must independently determine whether or not they have appropriate legal grounds to use the data, something they frequently lack the resources or expertise to do. 
In practice, compliance is hard to enforce, and while datasets are increasingly accompanied by datasheets~\cite{GebruMorgensternEtAl_2020_Datasheets_for_Datasets} and similar documentation, navigating the legal structures involved is not always straightforward.
Public repositories do present a unique opportunity to help harmonize emerging standards around documentation~\citep{mcmillan-major-etal-2021-reusable} mentioned in Section~\ref{sec:context-technical}, but they are structurally unable to support the oversight and management that are essential to our purposes. Our values of autonomy, consent, and contestation are difficult if not practically impossible for public dataset repositories, due to the full reliance on self-governance by dataset submitters (but see the Wikimedia model for related mechanisms for content curation, Section~\ref{sec:current-approaches-wikimedia}).\vspace*{3pt}

\noindent\textbf{Open Data Initiatives.} 
In NLP, open data initiatives involve collecting, processing, and sharing data that is public, but inaccessible or difficult to use \citep{pile}.
Some prominent open data initiatives have developed in response to the practice at many companies of training ML models on unreleased data, including OpenWebText \citep{Gokaslan2019OpenWeb}, which seeks to replicate the dataset that GPT-2~\cite{radford2019language} was trained on; BookCorpus2 \citep{pile} and Smashwords21 \citep{retrospective-bookcorpus}, which seek to replicate the formerly public BookCorpus dataset \citep{BookCorpus}; and LAION-400M \citep{LAION-400M} which seeks to replicate the WebImageText dataset that CLIP and DALL-E were trained with~\citep{Radford2021LearningTV}.
Another form of data replication effort seeks to provide public access to previously privately held data.
C4, the dataset that the T5 language model was trained on \citep{raffel2019exploring}, went unreleased for almost two years until it was replicated and shared by the Allen AI institute, enabling other scholars to study it and use it for training their own models.
Open data initiatives meet many of our desiderata for data governance, but possess some key omissions.
Critically, the goals of reproducible research that underlie the public recording of datasets are inherently in tension with the need to update datasets to accommodate requests to remove personal information \cite{rogers2021just}, and unredacted copies may circulate for years~\citep{Corry2021ThePO}.\vspace*{3pt}

\noindent\textbf{Example: Distributed Data Governance in the Wikimedia Project.}
\label{sec:current-approaches-wikimedia}

The Wikimedia projects offer a wealth of experience in highly collaborative and largely self-regulated data curation \citep{forte2009decentralization}, similar to the goals in the proposed governance structure.
The core stakeholders map to the proposed governance structure in Figure~\ref{fig:data-gov-org} as follows: the many contributors to the knowledge that is gathered on Wikimedia projects (data rights holders), editors (data custodians), the Wikimedia Foundation (data stewards and helpers), and the researchers, digital platforms, and many additional end-users of Wikimedia content (data modelers).
The Wikimedia projects face many of the same tensions that would face governance of global digital language data, such as diverse needs and goals of editors \citep{rader2020why}, the need to navigate varying local laws such as ``freedom of panorama'' \citep{commons_panorama} when determining whether an image can be hosted \citep{de2017public}, and how they are situated within existing power imbalances \citep{vrana2020toward}.

For example, to create some consistency for editors and end-users of Wikimedia data, the data is governed in part through content licenses.
Content licenses vary in attribution requirements between projects, but restrict contributors' rights on how their work is used.
This can be in conflict with cultural values, e.g., in the case of indigenous communities that are generally underrepresented on Wikipedia but have concerns about how their knowledge might be exploited if shared \citep{casemajor2019openness,carroll2020care}.
To ensure that the content adheres to the chosen licenses (and other regulations \citep{wiki:pillars}), editors have written policies (\ul{norms} as in Section \ref{sec:governance-values}) that are constantly evolving and being contested themselves \citep{butler2008don,wiki:unreliable}.
Similar to the proposed DSO, the success of the Wikimedia editor community is facilitated by a large ecosystem of \ul{tools} \citep{muller2013work,geiger2010work} (as in Section \ref{sec:governance-scope}) such as APIs, dumps, database replicas, and various cloud environments that can be used by tool developers to provide local access to this data~\cite{wiki:cloud}.
The ability for the community to build the tools required for data governance has been crucial to their success at scale \citep{halfaker2020ores}.

\section{A New Data Governance Structure}
\label{sec:dso}

Let us now review the needs we have identified for a governance structure in Sections~\ref{sec:governance}, \ref{sec:context}, and \ref{sec:current-approaches}.
We want an organization driven by a set of guiding \ul{values} outlined in Table~\ref{tab:bigscience_initial_values}, and notably the \ul{inclusion \& representation} of all categories of stakeholders identified in Table~\ref{tab:data_stakeholders}, in a fashion that fosters equitable access across social, cultural and geographical contexts (Section~\ref{sec:context-social}).
In so doing, the governance structure needs to account for the complexity and diversity of corresponding legal contexts (\ref{sec:context-legal}). 
We reviewed some issues and promising directions around the culture of data use in ML (\ref{sec:context-technical}) and current approaches to data management in the field (\ref{sec:current-approaches}), and found coordination across stakeholders following the desiderata detailed above to be a particular challenge.

The need to collect, share, access, and define norms, management, policies, guidelines, and values around the use of data suggests a structure with multiple categories of distributed actors prioritizing different aspects and communicating with one another for alignment on end goals, legal issues, values, and interoperability. 
To that end, we propose a data governance structure with six main actors, whose roles and relationships are summarized in Figure~\ref{fig:data-gov-org}.
The actors additionally interact with Data Sourcing and Data Tooling, as discussed in Section \ref{sec:governance-scope}.
In this Section, we start by describing the specific roles of the \textbf{data governance entities} involved in this structure.
We then review the relationships between these entities through two lenses: the \textbf{journey of the data} through the structure from its initial creators to the data modelers and the \textbf{role of the DSO} in formalizing frameworks and aggregating feedback and expressions of the various stakeholders' needs, especially with the aim of fostering the values in Table~\ref{tab:bigscience_initial_values}.

\begin{table*}
\small
\begin{tabular}{@{}l|p{0.6\textwidth}}
\hline
\textbf{Data Rights-holders} & Decide whether to share their data. \\\hline
\textbf{Data Providers} & Make data available to others. \\\hline
\textbf{Data Hosts} & Gather and hold data aligned to constraints  \\
& defined within the governance structure. \\\hline
\textbf{Data Modelers} & Specify dataset values and requirements. \\\hline
\textbf{Data Stewardship Org.} & Discussion space for all actors involved.\\\hline
\textbf{Data Helpers} & Ensure decisions respect rights and regulations \\ 
\hline
\end{tabular}
\caption{Actors within the proposed Data Governance structure}
\label{tab:gov_actors}
\end{table*}

\begin{table}
\small
    \centering
    \begin{tabular}{@{}l|l@{}}
    \hline
\textbf{Data Modelers +}  & Data dissemination \\
\textbf{Data Hosts} & Specific use case restrictions \\\hline
\textbf{Data Hosts +} & Data dissemination \\
\textbf{Data Providers} & Conditions for serving data \\
& Rights with respect to\\ 
& derived products    \\
\hline 
\end{tabular}
\caption{Binding agreements needed in the Data Governance Structure.}
\label{tab:governance_agreements}
\vspace*{-10pt}
\end{table}

\vspace*{3pt}

\noindent\textbf{Data Governance Entities}
Table~\ref{tab:gov_actors} summarizes the roles around which we organize the governance structure; specific entities may take \textbf{one or more} of these roles at various times (e.g., a data modeler may also make their own dataset and entrust it back to a data host as data provider). Figure~\ref{fig:data-gov-org} maps some of these roles to traditional categorization of data governance, including data steward and data custodian. We review each of these roles next.

Our effort toward defining data governance roles starts with asking \textit{where the data is found}, and \textit{whose rights} need to be accounted for.
\textbf{Data Rights-holders} are varied: they can be individuals, organizations or companies. An individual who wrote on social media, for example, might have legally protected \ul{privacy rights} on their language data used in a dataset (Section~\ref{sec:context-legal}), and organizations such as radio stations, newspapers, or content platforms have \ul{property rights} on the data they create or host. In general, the Data Rights-holders correspond to the \emph{data subjects} and \emph{data creators} categories of stakeholders in Table~\ref{tab:data_stakeholders} and represent the focus of the values of \ul{contestation}, \ul{consent}, \ul{privacy}, \ul{attribution}, and \ul{just rewards} described in Table \ref{tab:bigscience_initial_values}.
In particular, Rights-holders can inform how specific items of their language data may be used, in accordance with legal protections and values. 

Data is brought into our proposed governance structure by \textbf{Data Providers}. Companies that host or create language data can act as Data Providers, as can research organizations that create datasets from public or private data or archival institutions that work on preserving online or offline content (\textit{e.g.}, the Internet Archive).
The Data Providers can be identical to or separate from the Data Rights-holders, and can either fully specify what the data they bring into the governance structure may be used for, or specify it to the extent permitted by the original rights-holders.

Data is served by \textbf{Data Hosts} who gather and hold data from the Data Providers so as to meet the goals of the governance project and comply with legal requirements. This data is in turn made available to \textbf{Data Modelers}.
Data Hosts maintain their own, possibly post-processed version of the language data offered by data providers, and can decide which data they want to host (\textit{i.e.} they may decline to host some of the data offered by a data provider).
Depending on the jurisdiction of the Data Host and Data Provider, the Hosts may need a specific legal basis for holding certain kind of data or being eligible for some of the research exceptions outlined in Section~\ref{sec:context-legal}, which may go from being a nonprofit organization to having some form of public interest mission.
As outlined in Section~\ref{sec:governance-scope}, \ul{Data Sourcing} happens at the intersection between Data Providers and Data Hosts; the diversity and representativity of the data available to the governance organization will depend on the ability of the hosts to establish relationships and support the need of the greater variety of data providers.
Notably, this is easier when there is a degree of proximity between the hosts and providers so they have similar social and legal contexts \textemdash which motivates the need to have data hosts around the world to foster linguistic and cultural diversity of the available language data.
This proximity is also necessary to enacting meaningful \ul{contestation} rights at the Data Host level, as it will allow the requester (Data Rights-holders) and the enacter (Data Host) to share similar understandings of the notion and rely on a similar legal framework (as the extraterritorial applicability of data protection laws like the GDPR around the world is still an open question).

The last category of distributed actors of our proposed data governance organization are the \textbf{Data Modelers}, who can request access to data held by the Data Hosts to use according to the requirements set forth by the Data Rights-holders, Data Providers, and Data Hosts. The Data Modelers have their own data needs, including visibility of the data available across the data hosts, and ease of processing (e.g. through a unified format for all data sources).
Researchers may also need some degree of replicability for experiments run using the data held with the governance organization (see Section~\ref{sec:context-technical}), which needs to be understood in the context of the \ul{contestation} rights within the organization.

Finally, a \textbf{Data Stewardship Organization} provides a discussion space for all the above-mentioned actors and connections involved, communicating between Hosts, Providers, Data Advocates, and Data Rights-holders.
The DSO brings together representatives of all of the other roles, and is supported by \textbf{Data Helpers}, including lawyers and legal scholars representing all regions where the governance organization operates and advocacy groups focused on representing the interests of populations affected by data use and technology~\footnote{\href{https://d4bl.org/}{Data 4 Black Lives}, \href{https://www.odbproject.org/}{Our Data Bodies} are two such organizations in the US.}.
The role of the DSO in establishing and formalizing relationships between the actors is further outlined below, and it also serves as a central repository of technical tools, relevant documentations, and as a facilitator of interoperability.
Given this role, both the input of the Lawyers and of the Data Advocates is requested on new choices to account for both relevant regulations and their impact on the values listed in Table~\ref{tab:bigscience_initial_values}.

\vspace*{3pt}

\noindent\textbf{Journey of the Data.}
Before eventually being used for research or development of NLP system, the data follows a journey from its original creators, to the Data Providers who introduce it to the governance structure, to the Data Hosts that aggregate various sources, to the Data Modelers.
Each of these transfers and nodes in the path defines its own agreements between parties. In particular, these agreements are structured around the notion of \textbf{contractual flow-down}: each subsequent actor on this path is responsible for communicating the requirements and restrictions formulated by its predecessors in addition to its own.

In the exchanges \textbf{between Data Hosts and Data Providers},  Providers make data available, and if they have full rights on the data, they may specify use conditions. Some selection criteria might include research only, or use by organizations that meet certain criteria (such as non-profit status, or value statements). These restrictions may be \textit{explicitly} set down in a license agreement signed by the Host and the Provider (Table~\ref{tab:governance_agreements}). As an additional incentive for Providers to share their data, this license may require the Host to give the Provider access to any by-product of their data, such as analyses or processed versions (\textit{e.g.} with PII removed; see Section \ref{sec:governance-scope}). If the Provider is proposing data that is curated from external sources, especially text that is regulated by general data protection laws, there is also an \textit{implicit} relationship \textbf{between Data Host and Rights-holders}. In particular, the Data Host will be bound to honor \ul{contestation} requests when an individual finds that their private information, or data that they have a commercial right to, is included in data shared with the Host without their explicit consent. The Host would then be required to remove the particular data items from their dataset. This aspect requires Hosts to share data via access restrictions or binding agreements, as opposed to allowing copies to be freely downloaded and proliferated.

Finally, the exchange \textbf{between Data Hosts and Data Modelers} is bound by another set of licensing agreements, which need to reflect the restrictions flowing down from the Data Providers and other Rights-holders, any additional constraints expressed by the Data Host, and a non-dissemination clause. The Modeler may be required to obtain a fresh version of any dataset reflecting the most recent version of the data host after a fixed amount of time. The latter two are essential to ensure that data that has been deleted from a Host to answer a contestation request, or whose license with the Data Provider has expired, does not remain available.

\noindent\textbf{Role of the DSO.}
While the DSO itself is not a direct party in any of the agreements outlined above, its role is to facilitate interactions between all entities involved and assure interoperability between actors. For example, Data Providers might have reason to propose their own license, especially to support values that are misrepresented in legal frameworks relevant to them~\footnote{\href{https://www.temanararaunga.maori.nz/}{Maori Data Sovereignty License}}.
In such cases, respecting our value of \ul{inclusivity} and the \ul{local knowledge} of the Providers on how best to represent their community's interests means allowing them to use their own licensing in their interactions with Data Hosts rather then requiring them to use one designed by the DSO.
Conversely, some of the Data Providers, especially ones with fewer proper resources, might not have the legal expertise to develop their own data sharing licenses -- then, a DSO standard data license would foster \ul{inclusion}. We provide our proposal for such an agreement between the Data Hosts and Data Providers in Appendix~\ref{app:agreement}.

This tension reflects the governance trade-off between harmonization and independence mentioned in Section~\ref{sec:governance-concepts}. One additional complexity of allowing Hosts and Providers to use custom licenses arises when a Host aggregates data from several Providers to share with a Data Modeler. Without any categorization of the various Provider licenses, the Host would have to either develop a new custom license for each aggregation of data sources, or leave it to the Modeler to understand the interplay of the various constraints. 
We address these issues through a dual approach. First, the DSO provides a \text license template for use in exchanges between the Data Providers and Data Hosts.
Second, the DSO \textit{maintains a taxonomy of licenses} designed to support rules for aggregating use case restrictions from Providers for the agreement between the Hosts and Modelers, to be updated when a new license appears that is not easily categorized.\footnote{CLARIN (Section \ref{sec:current-approaches}) uses a restricted version of such a taxonomy for non-commercial data only~\citep{Kelli2018ImplementationOA}}

This approach exemplifies the general role of the Data Stewardship Organization at various points of exchange in the governance organization, in its two roles as a fallback mechanism or default option for actors that do not have the resources to develop their own processes, and as an enabler of interoperability between processes when they do: whether the process in question corresponds to the license agreement between the Data Host and Data Modelers, the framework for identifying a legitimate contestation request between a Data Host and Rights-holders, or the \ul{technical tools} used for identifying personal information or managing access to data.

\section{Discussion}

We have introduced an approach to data governance, grounded in an ongoing year-long case study that coordinates data internationally to train a Large Language Model. Critical aspects of the governance structure include protocols for achieving different values, working with established norms, and contending with the different laws applicable across datasets. This requires coordinating multiple stake-holders and rights-holders. 
Our approach is modular, where different parties focus on different aspects of the dataset processing and sharing, interconnecting data providers, data hosts, and data developers.  This is coordinated by a Data Stewardship Organization that develops appropriate management plans, access restrictions, and legal scholarship.
A complementary Data Tooling efforts help to provide resources common to the legal and ethical needs of the participating institutions,
We have found that one of the most difficult hurdles is developing legal agreements for Providers, Hosts, and Modelers that respect the laws and copyrights set forth in the data, as well as the laws of the institutions' regions.
We tackle this problem through the lens of stated governance \textit{values}, which inform the kinds of agreements that are necessary.

\bibliographystyle{ACM-Reference-Format}
\bibliography{lang_data_gov}

\onecolumn
\appendix

\section{Crafting Values in Data Governance}
\label{app:values}

\begin{figure}[h]
    \centering
    \fbox{
    \includegraphics[scale=0.3]{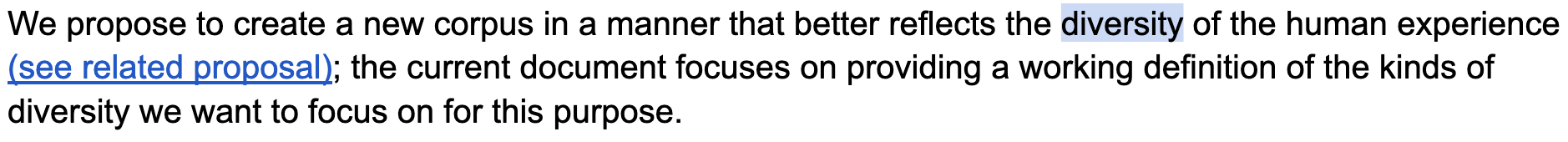}}
    \caption{Snippet from initial Data Governance planning doc, with \textit{diversity} value highlighted.}
    \label{fig:values_initial_annotation}
\end{figure}

\begin{figure}[h]
    \centering
    \fbox{\includegraphics[scale=0.3]{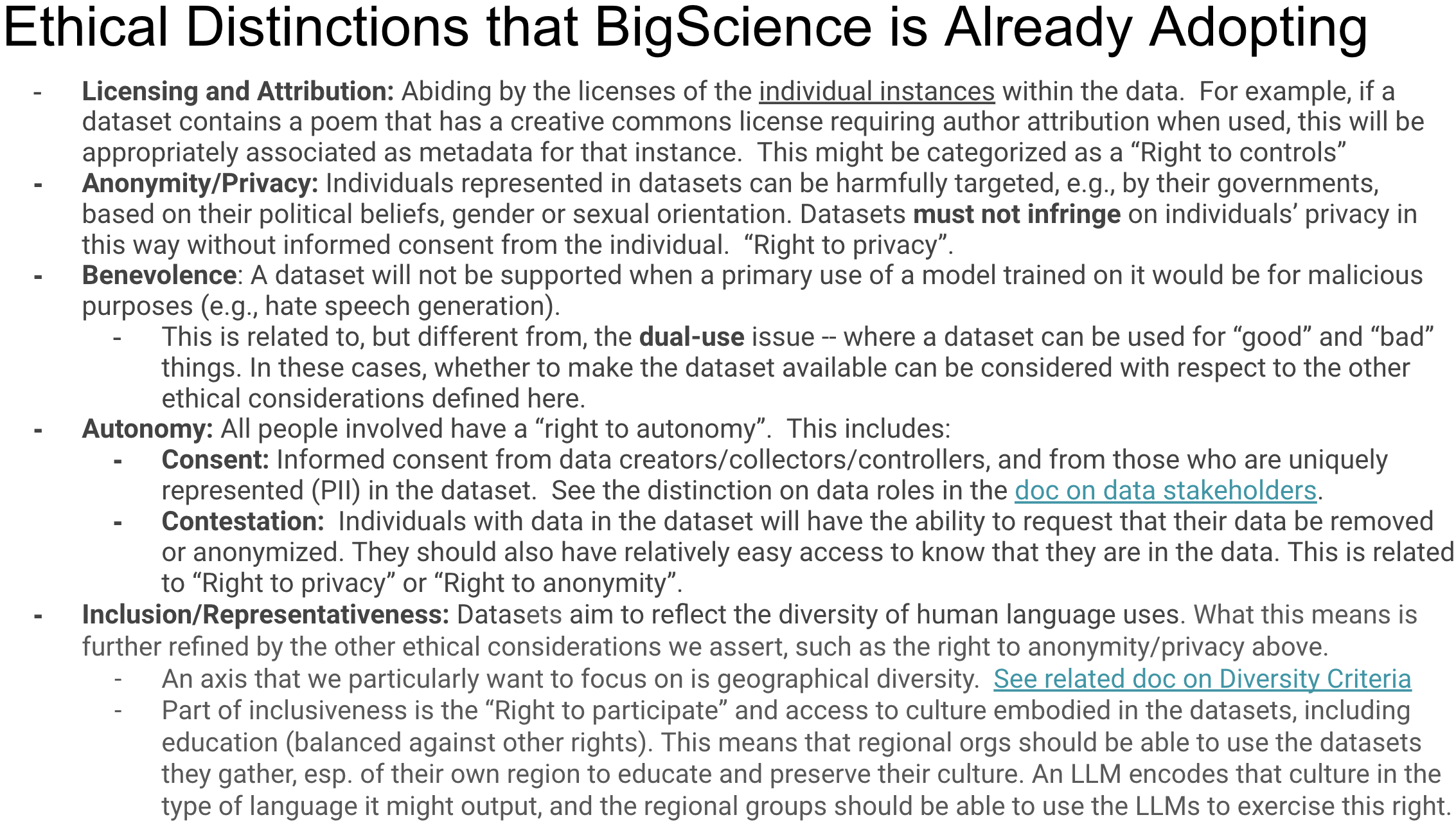}}
    \caption{Screenshot of the earliest draft of values and definitions discussed live.}
    \label{fig:values_iteration1}
\end{figure}

\begin{figure}[h]
    \centering
    \fbox{
    \includegraphics[scale=0.3]{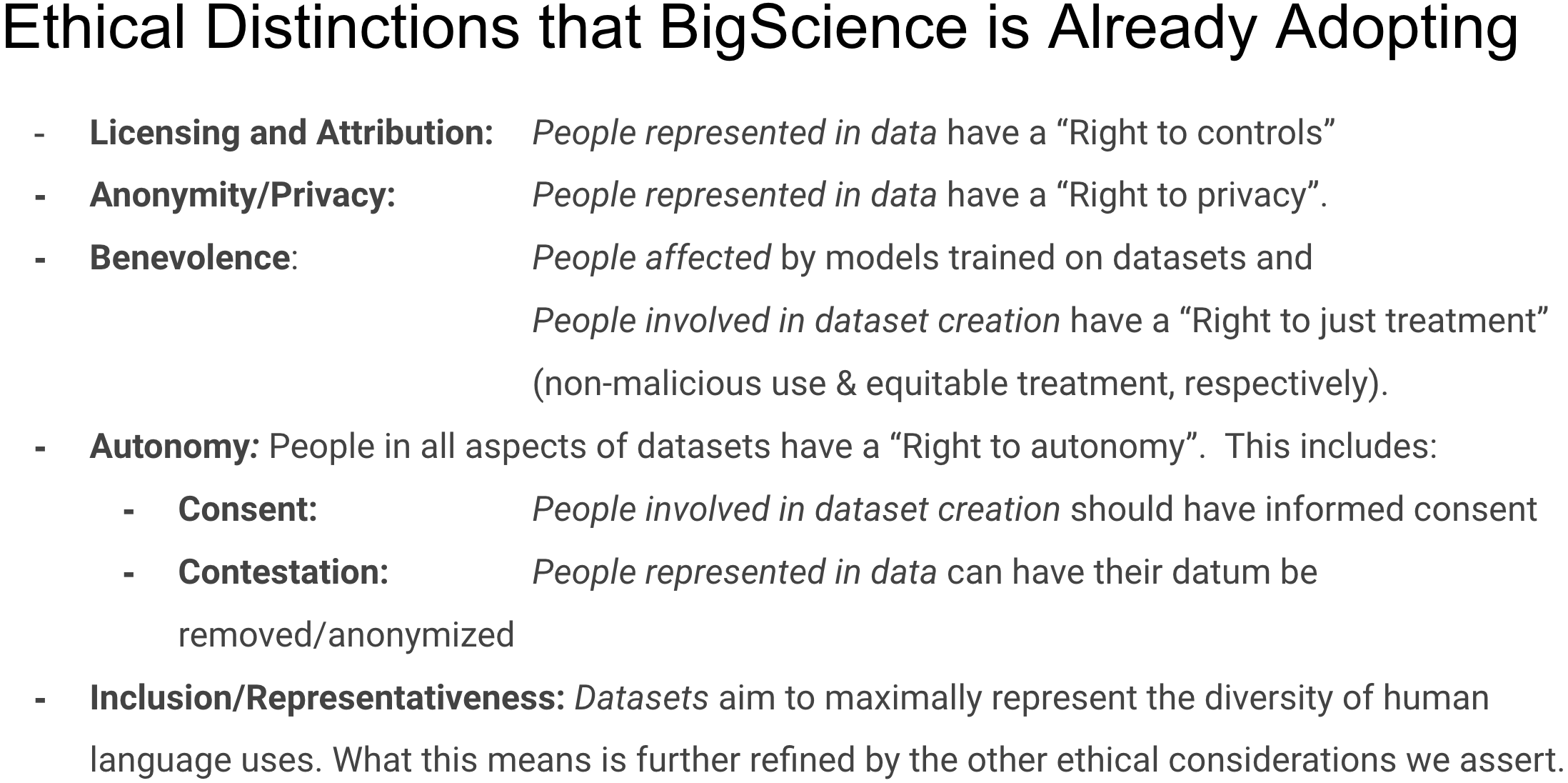}}
    \caption{Screenshot of an early draft of values and definitions discussed live.}
    \label{fig:values_iteration2}
\end{figure}

\pagebreak

\begin{figure}[h]
    \centering
    \fbox{
    \includegraphics[scale=0.3]{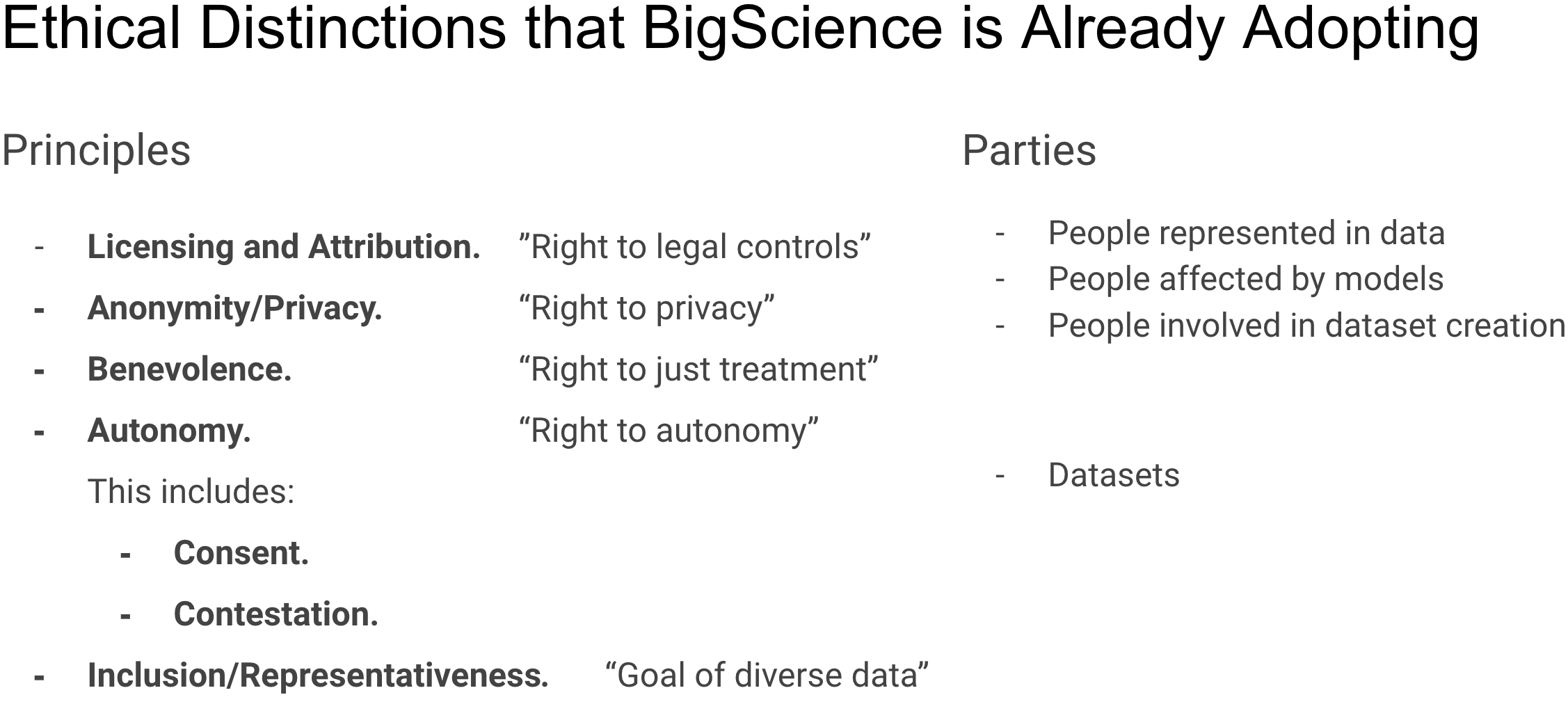}}
    \caption{Screenshot of a revised draft of values and definitions discussed live.}
    \label{fig:values_iteration3}
\end{figure}

\subsection{Overview of Approach}
An initial seed set of values for the data governance project were first implicitly expressed by the project planners, in our initial planning documents.  These documents were a product of roughly a handful of people: The group co-leads (3 people) who primarily authored them, and a set of people who provided comments and additions after the planning documents were shared more broadly within the BigScience effort.

Discussions and debates within the larger working group (around 10 people) at our regular meetings refined and expanded on these values in light of what everyone in the group wanted to prioritize in the project. For example, notes on needing to take care to make sure the data wasn't inappropriately reductive towards some populations was tentatively labelled as \textit{inclusion}, and with the working group, this eventually evolved into a value of \textit{representativeness}, defined as capturing the full diversity of human language use.

Similarly, the working group together decided on the best terms to use for the different value proposals. Our goal was to align on shared values to help prioritize different aspects of the work and to have some guidance to inform the decisions and potential disagreements we'd have as a working group moving forward. We recognized this was especially important as more people became involved, and so sought to have a basic set of values in place within the first 2 months of the project. Notably, prioritizing different aspects of \textit{inclusion} was a strongly shared goal across participants. 

\subsection{Steps}

To create the initial set of values, we first reflected on the fact that no one would be operating as a ``blank slate'' in this working group; that we all had our own values, and our own goals and motivations in working on the project.  As such, we focused on identifying what values we were \textit{already bringing to the table}. This was an exercise of making the implicit explicit, and required annotating the initial planning documents alongside larger working group discussions.

First, we organized all documents and notes for the initial creation of BigScience and the working group in chronological order.  Then, the co-lead went through each, highlighting specifically mentioned values -- such as geographical \textit{diversity} -- as well as annotating implicit values expressed in the text by the various authors and commenters. For the latter approach, the terms used for the annotations served as placeholders for further discussion within the larger working group. 

Then, the working group discussed the highlighted values and value annotations in light of their surrounding text, what the implicit ideas behind the text were, what we all felt we should be doing relevant to the value, and what we were all understanding and not understanding. Throughout these discussions, we crafted definitions live for what these value terms meant.  Once the definitions were in place and generally agreed upon, we discussed the specific terms used as value labels, and in some cases changed them, or broke up definitions into different components to identify more than one value.  

Screenshots representative of how these discussions evolved, in chronological order, are depicted in Figures \ref{fig:values_initial_annotation}, \ref{fig:values_iteration1}, \ref{fig:values_iteration2}, and \ref{fig:values_iteration3}

\subsection{Growing}
Over the course of the project, the size of the working group grew. From an initial set of around 10 people, we became a group of 50+ (some more involved than others), with some individuals taking on different roles as needs arose (for example, legal scholars and others interested worked on crafting a Data Host-Provider Agreement). All participants were introduced to the grounding values and the overall plan for the governance structure as they joined; indeed, presentations on this content arguably brought more people into the group.

\section{DSO Standard Host-Provider Agreement}
\label{app:agreement}

As outlined in Section~\ref{sec:dso}, one of the ways the DSO fulfills its purpose is by providing templates for licenses and leagal agreements between parties. The following license can be used to formalize a legal agreement between a \textbf{Data Host} and \textbf{Data Provider} in a way that supports our proposed governance values.\vspace*{-0pt}



\section*{Supplementary material}

# DATA PROVIDER-HOST AGREEMENT

*v0.1*

1. PREAMBLE

BigScience is an open research collaboration involving over 1000 participants from 60 countries, focusing its collaborative research efforts in the study and development of natural language processing systems (hereinafter NLP).

The project is motivated by recent evolutions in the field brought about by the growing capabilities, popularity, size and cost of Large Language Model-based methods. The computational resources and data needed to develop LLMs are affordable by a handful of institutions, who often conduct this research behind closed doors despite its significant impact on society.

Thanks to the support of a large compute grant on the French Jean Zay public super-computer, the participants of BigScience can instead collaborate across a range of academic institutions and organizations to create an openly accessible Large Language Model (LLM), available for the general public. This can be used to fuel research, governance, regulation, and future technology.

In particular, the choice and governance of the Data used to develop these technologies are of paramount importance. Previous work has mainly relied on text obtained from snapshots of the Internet, due to the large amount of Data and availability. Unfortunately, this convenience choice raises multiple ethical and legal issues and leads the technology to amplify harmful biases in its deployed applications.

BigScience takes an alternative approach of identifying Data sources for a training corpus. Namely, our participants built an annotated catalog of high-quality language resources to cover the diversity of languages and social contexts that should make up such a training corpus. There are two essential parties in charge of making this data available, under the auspices of BigScience: First, the Data Providers, any institution willing to license datasets of interest purely for research purposes on a royalty free basis; and Second, the Data Host, institutions willing to contribute their technical capabilities in order to host the data provided, enabling society to access it. These are the champions of data sharing and openness in research. This License governs the use of Data as informed by the BigScience Ethical Charter. BigScience has set forth its Ethical Charter representing the values of its community. Although the BigScience community does not aim to impose its values on potential users of the Data, it is determined to take tangible steps towards protecting the community from inappropriate uses of the work being developed by BigScience.

Consequently, the main objective of this Data Provider Agreement (the Agreement) is to serve as the core instrument enabling and governing the sharing of data between the interested parties, for the benefit of open research. Both parties strive to serve this goal by entering into this Agreement.

2. DEFINITIONS

**"Agreement"** means this Agreement including all its Exhibits.

"**Confidential Information**" means information that one Party discloses to the other Party under this Agreement and that is marked as confidential or would normally be considered confidential.

"**Data**" means machine-readable informational content (individually or as a whole i.e., collection of Datasets) made available by the Data Provider.

**"Meta-Data"** means supplementary information of the Data, for example, summaries or visualizations of the data, restricted excerpts, authorship information and high-level statistics (i.e. word counts)).

"**Dataset**" means one specific collection of Data on which the Data Provider has the necessary rights enabling the latter the sharing of it under this agreement.

**"Processed Dataset"** is a Dataset produced via Data transformations, including additional modifications to one dataset (including PI(I) removal, additional annotations, extracted text, subsetting by language, removal of individual data points), dataset combinations, etc.

"**Data Host**" means a legal entity permitted to process, prepare, and manage subsequent 3rd party access to the Data of the Data Provider under the scope of this agreement.

"**Data Provider**" means the individual or legal entity granting permission to the Data Host to access and further manage the Data for the purpose of this Agreement.

**"Derived Work"** means any artifact created using Data covered by this Agreement.

"**Parties**" means any individual or entity entering into this Agreement.

"**Third Parties**" means individuals or legal entities that are not controlled by any of the involved parties in this Agreement.

**"User"** means individual and/or legal entity having access to the data provided by the Data Provider and hosted by the Data Host for the purpose of this Agreement.

3. PURPOSE, RIGHTS GRANTED & SCOPE

The Data Provider grants to the Data Host a non-exclusive, non-transferable, non-sublicensable, irrevocable, perpetual, royalty-free and worldwide license to use (that is access, store, prepare, process, label and/or share) the agreed upon Data (see List of Datasets in Exhibit A) in accordance with the use case scenarios and further (re)distribution policy, as stated in Annex III (see below).

4. DATA PROVIDER RIGHTS AND OBLIGATIONS
    a. The Data Provider warrants that it is the owner of the Data or has the necessary rights to enter into this Agreement regarding the Data listed in the Dataset section (see Exhibit A).
    b. The Data Provider will provide Data Host with valid contact information in order to settle any queries or issues related to the Data.

c. The Data Provider will provide the Data Host access to the Data in a suitable format agreed upon by both parties.
d. The Data Provider shall not be subject to any damages or liabilities for any malfunction, error or omission in the Data. In case the Data Provider becomes aware of, it will diligently inform the Data Host in order to implement the proper modifications. From its side, the Data Host will do the same.
e. In case the Data Provider is informed about the application of restrictions of any kind the Data Provider will notify the Data Host. For instance, in case of becoming knowledgeable of any actual or suspected intellectual property rights infringement, damages or claims associated with the Dataset the Data Provider promptly notifies the Data Host such that the further infringing usage of the Dataset can be stopped.
f. The Data Provider acknowledges that the Data does not contain any malicious source-code that adversely affect, alter, damage or destroy the proper functioning of any software, operating system and/or hardware this may include but is not limited to viruses, trojan horses ransomware, back doors and spy software.
g. The Data Provider shall inform the Data Host in case the Data Provider becomes aware that the dataset does not comply with relevant regulations and laws, such as personal data-related regulations.
h. The Data Provider hereby disclaims any representations and warranties of any kind, express or implied, including without limitation any warranties of fitness for the purpose set out in this Agreement or beyond regarding the Data. The Data Provider does not guarantee the accuracy, adequacy or completeness of the Data.

5. DATA HOST RIGHTS AND OBLIGATIONS
    a. No rights are granted to the Data Host with respect to the Data other than those stipulated in this Agreement, except any exceptions or limitations provided by law.
    b. The Data Host will hold harmless the Data Provider against any claims, demands, suits or damages arising from the use of the Data in accordance with the purpose set out in this Agreement.
    c. The Data Host acknowledges and agrees that the following disclaimers apply to all End-Users and other entities who have access to the Data The Data is provided "as-is".
    d. In case of becoming knowledgeable of any actual or suspected intellectual property rights infringement, damages or claims associated with the Data the Data Host will promptly notify the Data Provider and stop the further usage of the Data until the issue in question can be resolved.
    e. The Data Host will not assert rights over any Data (excluding Meta-Data) made available by way of this Agreement.
    f. The Data Host will undertake commercially reasonable efforts to appropriately attribute the Data Provider as the source of the Data.
    g. The Data Host is allowed to create and publish research (including benchmarks, performance indicators and/or scientific insights) gained using the data under this agreement, for the purposes of BigScience's research scope.

h. The Data Host will undertake commercially reasonable efforts to remove personal data and information from the Dataset before using the Dataset Notwithstanding the latter undertaking, the Data Provider should, under Section 4(j) of this agreement, inform the Data Host in case the former is aware of the existence of personal data under the licensed dataset(s).

6. LIMITATIONS
   a. This agreement grants access to the Data exclusively for the purpose and chosen Data Access Policy (see Exhibit A) stated in this Agreement and does not extend to any other purpose nor does it apply to any other data not listed in the Dataset section (see Exhibit A).
   b. If the Data Provider complies to the data management plan the Data Provider holds the Data Host free and harmless of any action, recourse or claims made by any third party due to the non-observance by the Data Provider of its obligations under this Agreement and intellectual property and/or personal data related 3rd party claims.
   c. Both Data Provider and Data Host will not be liable for any processing activities of the Data under this agreement by any User having access to it under the framework of BigScience.

7. FEES AND COSTS

Neither party will charge any fees, royalties or costs associated with implementing this Agreement. All accruing costs or expenses of any party in relation to this Agreement are solely to be carried by the responsible party alone.

8. SECURITY

The Data Provider shall make reasonable efforts to provide the Data to the Data Host using up-to-date security standards (this may include but is not limited to data transmission via secure transport protocols, storage on secured servers as well as secure data processing). In case the data is made accessible via authentication the Data Host ensures that the used authentication method meets up-to-date standards.

9. TERM AND TERMINATION
   a. This Agreement is valid from the date the involved parties agree, or by default, from the moment it is signed by all the involved parties.
   b. The term of this Agreement shall be from the Agreement Date until the last to expire of the Data Provider's intellectual property rights or any related rights on the licensed Dataset, strictly for the purpose of this Agreement.
   c. This Agreement can be terminated by either party immediately in case the other party breaches this Agreement upon due notice of it and the breach is not remediated within 30 days.
   d. In case either party would like to voluntarily terminate the Agreement, it shall act in good faith and provide the other party with (i) a reasoned statement justifying the decision; (ii) a 60 days pre-advice; (iii) and, give the other party the opportunity to negotiate any new terms and conditions for the sake of the Agreement's continuity, and beyond, for the sake of BigScience's research goals.

e. Upon termination of this Agreement for any reason, the Data Host and Data Provider cease to use the Data and Processed Datasets and for the purposes set out in this Agreement within 14 days and upon that delete the Data and Processed Datasets immediately (respecting any holding periods or processing information storage required by the law) . This does not affect already completed or created Derived Work before the termination of this Agreement.

10. FORCE MAJEURE

Neither party shall be liable to the other for a failure of performance undertaken in this Agreement if prevented from doing so by any circumstances beyond its reasonable control (such as but not limited to fire, flood, drought, war, explosion, terrorism, computer hacking and viruses, acts of any government body, perils of the sea and air).

11. CONFIDENTIALITY

Each party shall treat this Agreement and all information and/or business practices of the other party it aquires or becomes knowledgeable of as confidential. Confidential information does not include any public or generally available information or any information independently obtained or available prior to entering this Agreement. Notwithstanding the foregoing, either party is allowed to reveal confidential information if it is required by law to do so.

12. ENTIRE AGREEMENT

This Agreement including its exhibits and attachments constitute the entirety of the Agreement between the parties and supersedes any prior negotiations or understanding.

13. MODIFICATION AND AMENDMENT

This Agreement can be amended or modified by mutual consent at any time. The amendment and/or modification must be put forth in writing.

14. DISPUTE RESOLUTION

Any dispute that may arise from the breach of this Agreement will be first subject to an alternative dispute resolution phase under the auspices of the BigScience Community.

15. SURVIVAL

The provisions set forth in section 6(b) (Limits of Liability), 10 (Term and Termination), 12 (Confidentiality), 15 (Governing Law), 16 (Survival) and Exhibit A (Section Restrictions of Use in the Dataset section) shall survive the termination of this agreement and continue to bind both parties.

16. SEVERABILITY

If any provision of this Agreement is held to be invalid, illegal or unenforceable, the remaining provisions shall be unaffected thereby and remain valid as if such provision had not been set forth herein. The parties agree to substitute such a provision with a valid provision most closely resembling the intent of such severed provision.

17. NO ADDITIONAL TERMS

Unless and to the extent expressly agreed to in writing between the Data Host and the Data Provider no other terms and conditions shall be binding to either party.

18.     FULL UNDERSTANDING
The parties acknowledge that they fully understand and agree to all of their rights and obligations under this Agreement.

DATA PROVIDER                          DATA HOST

……………………………….                      ……………………………….
Name                                   Name

……………………………….                      ……………………………….
Date and Location                      Date and Location

# Annex

**DATA PROVIDER SCHEDULE (EXHIBIT A)**

    I.    Data Provider Information

|  |
|---|
|  |

Data Provider Name

|  |
|---|
|  |

Data Address

|  |
|---|
|  |

Data Provider Contact Information

|  |
|---|
|  |

Data Provider Name

|  |
|---|
|  |

Special Conditions (*if applicable*)

|  |
|---|
|  |

Data Management Plan

    II.    Datasets

|  |
|---|
|  |

List of Datasets

|  |
|---|
|  |

License of Datasets (*if more than one, please assign in List of Datasets*)

|  |
|---|
|  |

Restrictions - Please indicate of any restrictions apply to any of the above listed datasets

III.    Field of Use

Scope / use cases:

    ☐ under condition: openly released models, results, and artifacts
    ☐ under condition: use RAIL license for ML artifacts (has to be attached)
    ☐ under condition: value alignment (determined by data host)
    ☐ under condition: value alignment (data modelers sign click-through form)

IV.    Data Distribution Policy

Acknowledging the immense value and benefits that your datasets may provide, and being conscious and respectful towards the different economic interests that you may have, this Agreement offers the Data Provider a flexible set of optional frameworks for the use, re-use, and distribution of data:

    ☐ The Data Provider permits the Data Host to use the Data for the purpose set out in this Agreement. The Data Host is **not** allowed to make the Data publicly available outside of the remits of this license (this does not include Meta Data).

    ☐ The Data Provider permits the Data Host to make the Data (as a whole or in parts or processed) available to downstream users upon signing a non-dissemination agreement.

    ☐ The Data Provider permits the Data Host to make the Data (as a whole or in parts or processed) available to downstream users using a system that supports authentication/synchronization

    ☐ The Data Provider permits the Data Host to make the Data (as a whole or in parts or processed) available with modifications such as anonymizing personal and/or sensitive information about individuals.

☐ The Data Provider permits the Data Host to use the Data for the purpose set out in this Agreement. Additionally, the Data Host is allowed to make the Data publicly available under the Data license (select one) provided by the Data Provider.

☐ CC BY 4.0 (Link)
☐ CC BY-NC-ND 4.0 (Link)
☐ CC BY-NC-SA 3.0 (Link)
☐ CC BY-NC-SA 4.0 (Link)
☐ CC BY-SA 3.0 (Link)
☐ CC BY-SA 4.0 (Link)
☐ CC-BY-NC 4.0 (Link)
☐ Microsoft Research Data License Agreement (Link)
☐ custom license agreement (see Attachment if applicable)
☐ Linux Foundation CDLA Permissive
☐ Linux Foundation CDLA Restrictive



\end{document}